\documentclass{PoS}

\newcommand{\etal}{et~al.}
\title{VLBI Polarisation with the Yebes (EVN) and Hobart (LBA) antennae}

\ShortTitle{VLBI Polarisation with the Yebes and Hobart antennae}

\author{\speaker{Richard Dodson}\\
        Observatorio Astronom\'ico Nacional, Espa\~na\\
        and University of Western Australia\\
        E-mail: \email{richard.dodson@uwa.edu.au}}

\abstract{Work has been on-going for the development of the required code for
full polarisation processing of VLBI data using some new antennae
mounts. The extensions of AIPS allows the support of two new mount
types; the left-handed and right-handed Nasmyth antennae (Pico Veleta
in the GMVA and Yebes-40m in the EVN) and the EW-mount (Hobart in the
LBA). The data handling process is seamless, once the correct mount
type has been selected. All subsequent calls to the parallactic angle
subroutine PARANG will return the feed angles for Left or Right
Nasmyth or EW-mount. These are required, respectively, for Pico
Veleta, Yebes-40m (low frequency branch) and Hobart.}

\FullConference{The 9th European VLBI Network Symposium on The role 
of VLBI in the Golden Age for Radio Astronomy and EVN Users Meeting\\
		 September 23-26, 2008\\
		 Bologna, Italy}

\begin{document}

\subsection*{Introduction}
\vspace*{-0.3cm}
To solve for the VLBI polarisation calibration terms one needs to
remove the rotation of the feed. Until now the only feed terms
supported in AIPS, and there could be solved for, were the Cassegrain
and Equatorial types. The Giant mm-VLBI array (GMVA), the European
VLBI network (EVN) and the Australian Long Baseline Array (LBA)
included feed types which were not supported in AIPS; that is the
Nasmyth (Pico Veleta and Yebes) and EW-mount (Hobart) types. New
additions to the AIPS code, now included in the general distribution,
will allow the full polarisation calibration of these antennae, and
their respective arrays. These routines have been used to calibrate
the LBA and to produce the first polarised VLBI images of Methanol Masers
(Dodson 2008). Followup EVN and LBA observations are being made for
this project. Preliminary results from EVN network experiment N08K2
show successful feed angle calibration. Here we present recent results
from test-time LBA experiment VX014, showing the comparison of the
polarised flux seen by these observations and those of the MOJAVE survey
(with the VLBA).

\subsection*{Yebes first VLBI fringes with the EVN}
\vspace*{-0.3cm}
As part of the first VLBI light of the new 40m antenna at Yebes,
Spain, observations were performed at 22-GHz in the network monitoring
experiment N08K2. Fringes were obtained between Yebes and the other
antennae. In Figure 1 
we present the phase difference between the RCP and LCP data after the
feed angle phases have been corrected for, showing that this was
successful. It was not possible to take this analysis further, as the
amplitude calibration was poor, which makes the seperation of the
contributions from source polarisation and instrument polarisation
difficult.

\begin{figure}
\begin{center}
\includegraphics[width=.8\textwidth]
{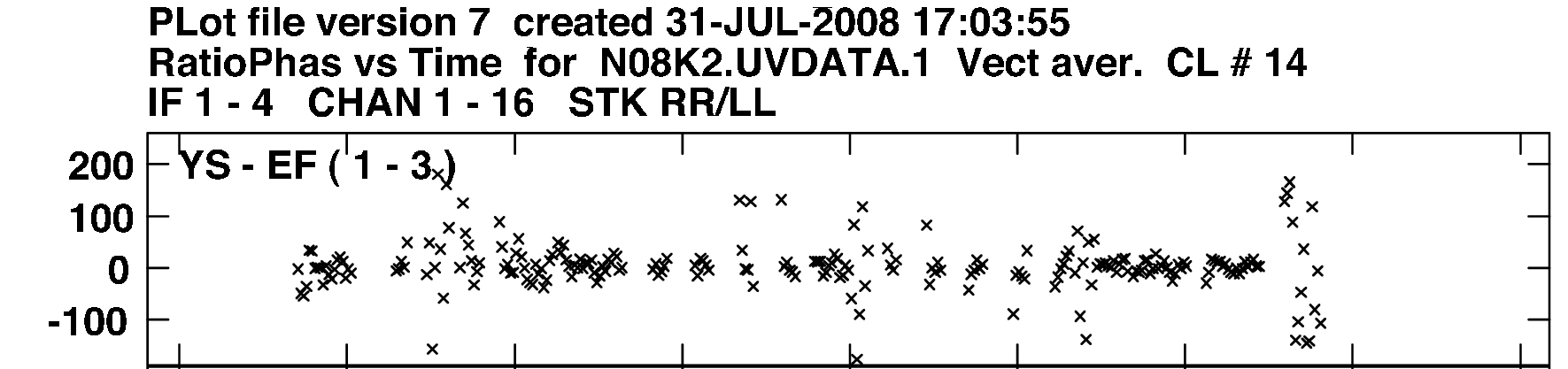}
\caption{RR/LL Phase between Yebes and Effelsberg after feed rotation correction.}
\label{fig:yb}
\end{center}
\end{figure}



\subsection*{Comparisons of the LBA with ATCA}
\vspace*{-0.3cm}
We have used the new code to successfully map the polarisation
direction of the magnetic fields of the Methanol Masers in
G339.88-1.26 (experiment V148).
Methanol Masers, it has been claimed, lend important support to the
model of the formation of Massive Stars from in-falling disks (Norris
\etal , 1993). The magnetic fields found via VLBI are not consistent
with the masers being in the disks, and therefore remove this strong
support for the disk model. 
%
%
As part of this analysis we compared the integrated spectral line
fluxes from ATCA and LBA data which showed agreement between the ATCA
detection of the polarised flux by velocity channel and that found 
with the LBA. See Dodson (2008) for the figures and details of the analysis.

\subsection*{Comparisons of the LBA with VLBA}
\vspace*{-0.3cm}
Four sources were observed in VX014, of which two are in the MOJAVE
program (Lister \etal , 2005); the Monitoring Of Jets and AGNs with
VLBA Experiments. We compared the images of these two sources, 3C273
and 3C279, despite the differences in resolution and frequency between
the two observations. The LBA at 8.4-GHz has a resolution of
$\sim$3\,mas and the VLBA at 15-GHz has a resolution of
$\sim$0.5\,mas. Nevertheless we smoothed the VLBA images to the
resolution of the LBA, and amplitude scaled the LBA data to match the
fluxes; assuming a simple spectral index between the two
frequencies. With these simple corrections the total power images from
the two instruments are in very good agreement, giving us the
confidence to compare the polarisation.

The images in Figure 2
show the total power, with a log scaled colour index. Overlaid are the
contours from the linear polarised flux for LBA and VLBA data. The
vectors are the field directions for the LBA data, these are in good
agreement with those of the VLBA data. However this is not total
independent, as the VLBA image was used to provide the calibration in
LPCAL. The D-term solutions for the two sources (both assuming a
polarised source and an unpolarised source) were in agreement to a few
percent, dominated by the errors in Mopra.
It is notable to the first order that there is a good match between
the results from the LBA and the VLBA, but at the second order there
are notable differences. These will need further investigation, as it
is possible that they are due to the second order effects from the
very high polarisation feed correction in the LBA calibration
terms. Work is undergoing to improve these, particularly for Mopra,
which were known to be anomalous.

\begin{figure}
\begin{center}
\includegraphics[width=.7\textwidth]{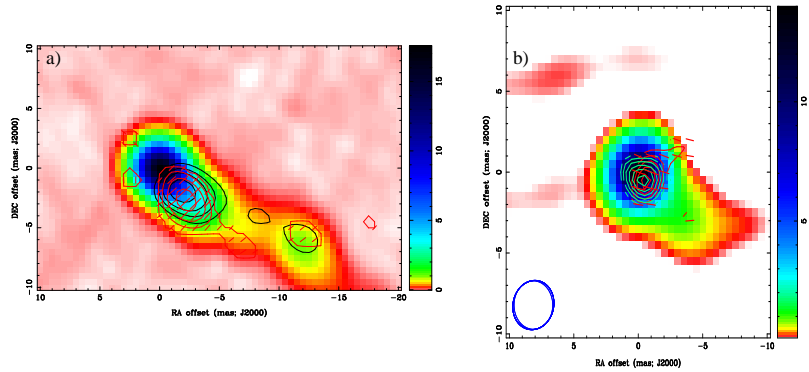}
 \caption
{a) The polarised flux from 3C273, as observed by the LBA and
   the VLBA in 2008. The colour scale is the LBA total intensity, the
   black contours are the VLBA polarised intensity, the red those of
   the LBA with the LBA polarisation vectors overlaid. b) The polarised
   flux from 3C279, as observed by the LBA and the VLBA in 2008. The
   colour scale is the LBA total intensity, the blue contours are the
   VLBA polarised intensity, the red those of the LBA with the LBA
   polarisation vectors overlaid.}
\end{center}
\end{figure}


\begin{footnotesize}

\end{footnotesize}

\end{document}